\documentstyle[preprint,revtex]{aps}
\begin{document}

\def\Db{\overline{D}}
\def\nle{{\stackrel{<}{\sim}}}
\def\nge{{\stackrel{>}{\sim}}}
\def\st{\widetilde{t}}
\def\stl{\st_{1}}
\def\sth{\st_{2}}
\def\mstl{m_{\stl}}
\def\mt{m_{t}}
\def\mz{m_{Z}}
\def\mw{m_{W}}
\def\tht{\theta_{t}}
\def\tew{\theta_{W}}
\def\gev{{\rm GeV}}
\def\slepton{\widetilde \ell}
\def\squark{\widetilde q}
\def\msq{m_{\squark}}
\def\msl{m_{\slepton}}
\def\photino{\widetilde \gamma}
\def\sfermion{\widetilde f}
\def\sneutrino{\widetilde \nu}
\def\selectron{\widetilde e}
\def\gluino{\widetilde g}
\def\ddf{{\rm d}}
\def\lam{\lambda'_{131}}
\def\zino{{\widetilde{Z}}}
\def\wino{{\widetilde{W}}}
\def\cbar{\overline{c}}
\def\sz1{{\widetilde{Z}}_{1}}
\def\swl{{\widetilde{W}}_{1}}
\def\ino{\widetilde{\chi}}
\def\misEt{\slash\hspace{-9pt}E_{T}}
\def\misP{\slash\hspace{-8pt}P_{T}}
\def\tanbe{\tan\beta}
\def\ccbar{\overline{c}}
\def\lam{\lambda'_{131}}
\def\frb{F_{\rm RB}}
\def\i{{\rm i}}
\def\xq{(x, Q^{2})}
\def\tui{T^{u}_{\i}}
\def\tuib{\bar{T}^{u}_{\i}}
\def\tdi{T^{d}_{\i}}
\def\tdib{\bar{T}^{d}_{\i}}
\def\tuo{T^{u}_{1}}
\def\tuob{\bar{T}^{u}_{1}}
\def\tdo{T^{d}_{1}}
\def\tdob{\bar{T}^{d}_{1}}
\def\tut{T^{u}_{2}}
\def\tutb{\bar{T}^{u}_{2}}
\def\tdt{T^{d}_{2}}
\def\tdtb{\bar{T}^{d}_{2}}
\def\tdh{T^{d}_{3}}
\def\tdhb{\bar{T}^{d}_{3}}
\def\tdf{T^{d}_{4}}
\def\tdfb{\bar{T}^{d}_{4}}
\def\ee{e^{e}}
\def\eu{e^{u}}
\def\ed{e^{d}}
\def\mz{m_{Z}}
\def\ael{A^{e}_{L}}
\def\aer{A^{e}_{R}}
\def\adl{A^{d}_{L}}
\def\adr{A^{d}_{R}}
\def\aul{A^{u}_{L}}
\def\aur{A^{u}_{R}}
\def\qz{{{Q^2}\over{Q^2 +\mz^2}}}
\def\utg{(Q^2-sx-\mstl^{2})}
\def\stg{(sx-\mstl^{2})^{2}+\mstl^{2}\Gamma_{\stl}^{2}}

\draft
\preprint{ITP-SU-92/03}
\preprint{TMU-HEL-9307}
\begin{title}
Signals for Scalar Top Quark at $ep$ Collider HERA
\end{title}
\author{
    Tadashi Kon$^*$,
    Tetsuro Kobayashi,
    Shoichi Kitamura, \\
    Keiichiro Nakamura and Shunji Adachi\\
}
\begin{instit}
$^*$Faculty of Engineering, Seikei University, Tokyo 180, Japan \\
Department of Physics, Tokyo Metropolitan University, Tokyo 192-03,
 Japan \\
\end{instit}
\begin{abstract}
In the framework of the minimal supersymmetric standard model (MSSM)
and the R-parity breaking model (RBM),
we investigate production processes of the scalar top (stop) at HERA
energies. These models are characterized by the possible existence
of the light stop whose mass is lighter than those of the top quark
and the other squarks. It is shown that in the MSSM the stop pair
production via boson-gluon fusion gives a sizeable cross section and
the most serious background $ep\rightarrow ec\bar{c}X$ could be
suppressed by appropriate kinematical cuts. We also show that in
the RBM the stop is singly produced in the neutral current processes
and we have a clear signal as a sharp peak in the Bjorken parameter
$x$ distribution of the scattered electron.
\end{abstract}

\narrowtext

\section{Introduction}

  The standard model (SM) of strong and electroweak interactions
describes amazingly well particle physics phenomena up to
the electroweak breaking scale. Particularly, the agreement between
theory and experiment in the LEP experiments was so impressive.
Nevertheless, we know that the SM could not be the ultimate theory
of everything, so long as there are a large number of free
parameters, the arbitrariness of particle masses and mixing
angles, and the lack of any explanation for the replication of
generations and so on. Still there is no experimental sign of new
physics beyond the SM.

  Various theoretical attempts are open to go beyond the SM.
Among them the supersymmetry (SUSY) \cite{Nilles} seems to provide us
an elegant and reasonable approach to meet the expectation,
provided that the SUSY breaking scale is $O$(1TeV).
In the framework of SUSY, every boson (fermion) is accompanied
by a fermionic (bosonic) partner with equal mass and coupling.
The requirement of equal mass is relaxed by the broken SUSY
either spontaneously or explicitly through introducing
soft SUSY breaking terms like gaugino mass terms, scalar
mass tems or trilinear couplings of scalar particles.
Still, there is no experimental evidence for SUSY partners of any of
the known particles, and only some limits on their masses
are given \cite{massbnd}. The theoretical expectation for SUSY, however,
is getting stronger than ever.
The analysis of the running coupling constant by SUSY GUTs
recently done seems to encourage people who believe
in SUSY \cite{running}.

  The purpose of the present paper is to focus our attention on the
scalar top quark (stop $\stl$) and investigate its production
mechanisms and decay properties in the framework of the minimal
supersymmetric model (MSSM) and the R-parity breaking model
(RBM) \cite{BARGER}.
We calculate cross sections together with detectable signals
expected to observe at HERA, the world first electron-proton collider.
Contrary to predicted high mass of the top quark, which is not
discovered yet, the stop mass could be lower than that of the
top quark in a model \cite{HK}.
The expected mass of the stop is clearly within the reach of HERA.
This is one of our strong motivation of the present work.
Higher energies available than LEP and cleaner events than hadron
colliders are another reason which has driven us to HERA.
Particularly, the stop could be singly produced in $ep$ collisions
in the RBM \cite{stoprb,susana,Hewett} because of the existence of
the electron-quark-squark couplings.
Clearly HERA is the best machine to search for the stop in such models,
because we expect the remarkable peak structure in the Bjorken parameter
$x$ distribution.
The theoretical prediction of the discovery limit of the stop
depends upon various model parameters as well as the energy and
luminosity of HERA.
We search for a possible experimental signal
observable in $ep$ collisions.

  The present paper is organized as follows.
After the model building of the stop in
Sec.{\uppercase\expandafter{\romannumeral 2}} we discuss the
stop production in MSSM in Sec.{\uppercase\expandafter{\romannumeral 3}}.
Section {\uppercase\expandafter{\romannumeral 4}}
is devoted to the resonance production of the stop by invoking the
R-parity breaking.
Finally, in Sec.{\uppercase\expandafter{\romannumeral 5}}
we give concluding remarks.

\section{Model building}
In the framework of the supersymmetric standard model,
scalar fermion mass matrices are expressed by
\begin{equation}\renewcommand{\arraystretch}{1.3}
{\cal M}^{2}_{\sfermion}=\left(
                 \begin{array}{cc}
                   m^{2}_{\sfermion_{L}} & a_{f}m_{f} \\
                   a_{f}m_{f} & m^{2}_{\sfermion_{R}}
                 \end{array}
                \right),
\end{equation}
where $m_{\sfermion_{L, R}}$ and $a_{f}$ are the SUSY mass parameters
and $m_{f}$ denote the ordinary fermion masses.
We can see from Eq.~(1) that for the sleptons and the squarks
except for the stops, the left and right handed sfermions are mass
eigenstates in good approximation owing to small fermion masses
in the off-diagonal elements of the mass matrices.
On the other hand, the large mixing between the left and right
handed stops will be expected from the large top-quark mass
\cite{HK}, and the mass eigenstates are expressed by
\begin{equation}
\left({\stl\atop\sth}\right)=
\left(
{\st_{L}\,\cos\tht-\st_{R}\,\sin\tht}
\atop
{\st_{L}\,\sin\tht+\st_{R}\,\cos\tht}
\right),
\end{equation}
where $\tht$ denotes the mixing angle of stops :
\begin{mathletters}
\begin{eqnarray}
\sin 2\tht=
{\frac{2a_{t}\,m_{t}}
  {\sqrt{(m^{2}_{\st_{L}}-m^{2}_{\st_{R}})^{2}
          +4a^{2}_{t}\,m^{2}_{t}}}}, \\
\cos 2\tht=
{\frac{m^{2}_{\st_{L}}-m^{2}_{\st_{R}}}
  {\sqrt{(m^{2}_{\st_{L}}-m^{2}_{\st_{R}})^{2}
          +4a^{2}_{t}\,m^{2}_{t}}}}.
\label{sintht}
\end{eqnarray}
\end{mathletters}
We can easily calculate the mass eigenvalues of the stops as
\begin{equation}
m^{2}_{\stl\atop\sth}
         ={\frac{1}{2}}\left[ m^{2}_{\st_{L}}+m^{2}_{\st_{R}}
             \mp \left( (m^{2}_{\st_{L}}-m^{2}_{\st_{R}})^{2}
            +(2a_{t}m_{t})^{2}\right)^{1/2}\right].
\label{stopmass}
\end{equation}
The SUSY mass parameters $m_{\st_{L, R}}$ and $a_{t}$
are parametrized in the following way \cite{susy} :
\begin{eqnarray}
m^{2}_{\st_{L}}&=&M^{2}_{\st_{L}}
  +\mz^{2}\cos{2\beta}\left({\frac{1}{2}}-
{\frac{2}{3}}\sin^{2}\tew\right)+m^{2}_{t}, \\
m^{2}_{\st_{R}}&=&M^{2}_{\st_{R}}
  +{\frac{2}{3}}\mz^{2}\cos{2\beta}\sin^{2}\tew+m^{2}_{t}, \\
a_{t}&=&-(A_{t}+\mu\cot\beta),
\end {eqnarray}
where $M_{\st_{L,R}}$, $\tanbe$, $\mu$ and $A_{t}$ denote
the soft-breaking scalar masses, the ratio of two Higgs
vacuum expectation values ($=v_{2}/v_{1}$), the Higgs mass
parameter and
the trilinear coupling constant, respectively.
In Fig.~1 we show contours of the various lighter stop masses $\mstl$
in ($A_{t}$, $M_{\st}$) plane for given values of
($\mt$, $\mu$, $\tanbe$), where we take
$M_{\st_{L}}$$=$$M_{\st_{R}}$$=$$M_{\st}$.
We find that if SUSY mass parameters and the top
mass are the same order of magnitude, the cancellation could
occur in the expression for the lighter stop mass Eq.~(\ref{stopmass}).
So we get one light stop $\stl$ lighter than the top as well as
the other squarks for a wide range of the SUSY parameters.

After the mass diagonalization we can obtain the interaction Lagrangian
in terms of the mass eigenstate $\stl$.
In Fig.~2 we show the Feynman rules for the lighter stop $\stl$
couplings to the gauge bosons $\gamma$, $Z$ and $g$.
While the stop couplings to the gluon and photon do not depend on the
mixing angle $\tht$, the $Z$-boson coupling depends sensitively on $\tht$.
More specifically, it is proportional to
$C_{\stl}\equiv {\frac{2}{3}}\sin^{2}\tew - {\frac{1}{2}}\cos^{2}\tht$.
Note that for a special value of $\tht$$\sim$0.983,
the $Z$-boson coupling completely vanishes \cite{DH}.

\section{Stop production in MSSM}
  In the MSSM the light stop $\stl$ will be produced in pair via
boson-gluon fusion \cite{stopbg}:
\begin{equation}
e^{\pm}p \rightarrow e^{\pm}\stl\stl^{*}X.
\label{bgstop}
\end{equation}
Feynman diagrams for the sub-processes are depicted in Fig.~3.
Referring to Fig.~3 we find that the $\gamma -g$ contribution to the
sub-processes is much larger than that of $Z-g$ \cite{stopbg}.
Since the $\gamma -g$ contribution does not depend on the stop mixing
angle $\tht$, the total cross section is not sensitive to $\tht$, but
only depends on the stop mass $\mstl$.
Owing to the photon dominance, we can use
the Weizs\"acker-Williams approximation (WWA) ;
\begin{equation}
{\frac{\ddf\sigma}{\ddf z}}=\int\ddf y P(y)\int\ddf\eta
G(\eta, {\hat{s}'}) {\frac{\ddf{\hat{\sigma}}}{\ddf z}},
\end{equation}
where $y=(p\cdot q)/(p\cdot \ell_{e})$ and
$z=(p\cdot p_{f})/(p\cdot q)$ with the variables
defined in Fig.~3.
The WWA factorizes the cross section of the process
(\ref{bgstop}) into the probability for emitting photon from lepton ;
\begin{equation}
P(y)={\frac{\alpha}{2\pi}}{\frac{1+(1-y)^2}{y}}
\log{\frac{Q_{\rm max}^{2}}{Q_{\rm min}^{2}}},
\end{equation}
the boson-gluon fusion cross section involving the real photon
\begin{equation}
{\frac{\ddf{\hat{\sigma}}}{\ddf z}}=
{\frac{4}{9}}{\frac{\pi\alpha\alpha_{s}}
{{\hat{s}}^{'3}z^{2}(1-z)^{2}}}
\left[2\mstl^{4}-2\mstl^{2}{\hat{s}'}z(1-z)
+{\hat{s}}^{'2}z^{2}(1-z)^{2}\right]
\end{equation}
and the gluon distribution function $G(\eta, {\hat{s}'})$ \cite{DO},
where $\eta$ denotes the momentum fraction of the gluon in the
proton and ${\hat{s}'}=y\eta s$.
The numerical integration has been performed by using the
program package BASES \cite{bases}.
We have the total cross sections : $2.4$pb for $\mstl=30\gev$,
$0.5$pb for $\mstl=40\gev$, $5\times10^{-2}$pb for $\mstl=60\gev$
and $8\times 10^{-3}$pb for $\mstl=70\gev$.

Now, what is the experimental signature ?
How to suppress its background ?
Actually, the stop can decay into the various final states :
\begin{mathletters}
\begin{eqnarray}
\stl &\to& t\,\sz1   \\
     &\to& b\,\swl   \\
     &\to& b\,\ell\,\sneutrino \\
     &\to& b\,\nu\,\slepton \\
     &\to& b\,W\,\sz1 \\
     &\to& b\,\ell\,\nu\,\sz1 \\
     &\to& c\,\sz1,
\end{eqnarray}
\end{mathletters}
where $\sz1$, $\swl$, $\sneutrino$ and $\slepton$, respectively, denote
the lightest neutralino, the lighter chargino, the sneutrino and the
slepton.
If we consider the light stop with mass lighter than 40GeV,
the first five decay modes (12a) to (12e) are kinematically
forbidden due to the lower mass bounds for respective particles
;  $m_{t}$$\nge$90GeV, $m_{\sz1}$$\nge$20GeV,
$m_{\swl}$$\nge$45GeV, $m_{\slepton}$$\nge$45GeV and
$m_{\sneutrino}$$\nge$40GeV \cite{massbnd}.
So there left (12f) and (12g).
The one-loop mode $\stl\to c\,\sz1$ (12g) dominates over the
four-body mode $\stl\to b\,\ell\,\nu\,\sz1$ (12f),
because of the large logarithmic factor $\ln({M_{planck}}/{\mw})$
appearing in (12g) \cite{HK}.
So we can conclude that such light stop will decay into
the charm quark jet plus the missing momentum taken away
by the neutralino with  almost 100$\%$ branching ratio.

Naively, it will be expected that TEVATRON and/or LEP can set severe
bounds on the stop mass through the processes ;
$gg$ $\to$ $\stl\stl^{*}$ $\to $ $c\cbar\sz1\sz1$
(TEVATRON) and/or
$Z$ $\to$ $\stl\stl^{*}$ (LEP).
However, the situation is not so obvious.
Baer et al. \cite{Baer} have performed the analyses of the
experimental data of 4pb$^{-1}$ integrated luminosity
TEVATRON running, and have obtained the results that the stop could
easily be escaped the detection if $m_{\sz1}$ $\nge$ 10GeV.
Such large neutralino mass could make the charm quark jets softer.
Consequently the stop production cross section plotted against
the missing transverse energy becomes smaller than
the present upper bounds, where we impose cuts on the missing
transverse energy.
Moreover, we find that LEP cannot exclude the light stop
for appropriate mixing angle $\tht$.
In Fig.~4 we show the stop mass dependence of the decay width
$\Gamma(Z\to\stl\stl^{*})$.
The horizontal line corresponds to the present upper bound for
$\Delta\Gamma_{Z}$ at LEP \cite{LEP}.
We can find that the stop with mass $\mstl$ $\nle$ 20GeV has not been
excluded if the mixing angle $\tht$ is larger than about 0.6.
The origin of such sensitivity of $\Gamma(Z\to\stl\stl^{*})$
has already been discussed in the previous section.
Recently DELPHI has performed the analyses on the direct search
for the stop through $e^{+}e^{-}\rightarrow \stl\stl^{*}$
and also studied $\Delta\Gamma_{Z}$ \cite{LEP}.
The result is still negative.
According to their analyses we have
the excluded region in ($\tht$, $\mstl$) plane shown in Fig.~5.
In the same figure we show the excluded region at TRISTAN
assuming massless photino \cite{TRISTAN}.
In this figure we also show the typical searchable regions at HERA.
We can clearly see that HERA would cover the window near
$\tht\simeq$1 in LEP region.
The upper bound at HERA goes upward for large integrated luminosity
and efficient background suppression.

For the light stop ($\mstl\nle$40GeV) production at HERA,
the event signature is $ec\ccbar X$ with large missing momentum
due to unseen neutralinos.
The most serious background process comes from the direct charm
pair production $ep \rightarrow ec\ccbar X$ via boson-gluon fusion
process whose total cross section is about $1$$\mu$b without any cuts.
In order to suppress this background we find the following kinematical
cuts to be efficient ; 1)the lower transverse momentum $P_{D,\Db}^{T}$
cuts for both $D$ and $\Db$ mesons in laboratory (LAB) system,
2)the upper momentum transfer squared $Q^2$ cut and 3)the lower
missing transverse momentum $\misP$ cut in LAB system
4)the acoplanarity angle $\phi_{(D-\Db)}$ cut,
$\phi_{(D-\Db)}$ being the angle between the $D$-proton and
$\Db$-proton planes in the boson-proton center of mass system.
5)the polar angle $\theta_{D,\Db}$ cuts, $\theta_{D,\Db}$ being
the polar angle of $D$ or $\Db$ with respect to the initial
proton axis in LAB system.
In Fig.~6 we show the acoplanarity angle distribution for the stop
production followed by the decay
$\stl$$\rightarrow$$c\zino_{1}$$\to$ ($D$-meson) plus missing momentum
and those for the background with cuts mentioned above.
These events in Fig~6 were generated in the kinematical regions ;
5$<$$Q^2$$<$100(GeV/c)$^2$, 10$^{-3}$$<$$x$$<$1,
0.1 $<$ $|\sin\theta_{D,\Db}|$ and $\misP$ $>$ 5GeV/c
by making use of the generator AROMA \cite{aroma} with JETSET
\cite{JETSET}.
Here we took $\mstl=30$GeV, $m_{\zino_{1}}=18\gev$
and the integrated liminosity $L=50$pb$^{-1}$.
We see that the $D$-meson jets with $\cos\phi_{(D-\Db)}\nge 0$
can be used to separate the stop production from its background
coming from boson-gluon fusion process.

\section{Stop Resonances in RBM}
  In the RBM there could exist the exotic Yukawa couplings of the stop
$\stl$, which violate the R-parity defined by $(-)^{3B+L+2S}$.
Here $L$, $B$ and $S$ denote the lepton number, baryon number and spin,
respectively.
The R-parity breaking (RB) couplings in the supersymmetric models
\cite{BARGER} are required in order to explain the cosmic baryon number
violation, the origin of the masses and the magnetic moments of neutrinos
and some interesting rare processes in terms of the lepton and/or
baryon number violation.
Here we take a coupling of the stop $\stl$ ;
\begin{equation}
{\cal L}_{\rm int}=
{\lambda'_{131}}\cos\theta_{t}(\st_{1}\bar{d}P_{L}e
                         +\st_{1}^{*}\bar{e}P_{R}d)
\label{rbL}
\end{equation}
originated from the RB superpotential
$W=\lambda'_{ijk}L_{i}Q_{j}{\bar{D}}_{k}$, where $i\sim k$ are
generation indeces and $L$, $Q$ and ${\bar{D}}$ donote the
chiral superfields.
The couplings generated from the superpotential
are most suitable for the $ep$ collider experiments
at HERA because the squarks will be produced in the $s$-channel
in $e$-$q$ sub-processes.
Here we consider the coupling (\ref{rbL}), which generates
the $s$-channel stop $\stl$ production in the neutral current (NC)
process \cite{stoprb}:
\begin{equation}
e^{\pm}p \rightarrow (\stl X) \rightarrow e^{\pm}qX
\label{RBNC}
\end{equation}
and the relevant Feynman diagrams are depicted in Fig.~7.
The production of squarks in the first and second generation
at HERA has been discussed in Ref.~\cite{susana,Hewett}.

We calculate the inclusive differential cross section for the
NC processes $e^{\pm}p \rightarrow e^{\pm}qX$ with
polarized $e^{\pm}$ beams :
\begin{equation}
{{\ddf\sigma}\over{\ddf x\ddf Q^{2}}}[e^{\pm}_{L,R}]=
  {{2\pi\alpha^{2}}\over{x^{2}s^{2}}}
  \sum_{q}\left[q(x,Q^{2})\sum_{i}
T_{i}(e^{\pm}_{L,R}q)+
  \bar{q}(x,Q^{2})\sum_{i}
T_{i}(e^{\pm}_{L,R}\bar{q})\right],
\label{rbcr}
\end{equation}
where $x$ and $Q^2$ denote the Bjorken scaling parameter and
the four momentum transfer squared, respectively, and
$q(x,Q^{2})$ are the quark distribution functions \cite{DO}
in the nucleon.
The analytic expression for the coefficients $T_{i}(e^{\pm}_{L,R}q)$
is presented in the Appendix.
The cross section depends sensitively on the decay width of the stop.
In this calculation we assume BR($\stl\rightarrow ed$)$\simeq$100\%, i.e.,
\begin{equation}
\Gamma_{\stl}={{\alpha}\over{4}}F_{\rm RB}(\stl )\mstl,
\end{equation}
where the coupling strength $F_{\rm RB}(\stl )$
is defined as
\begin{equation}
F_{\rm RB}(\stl )={{{\lambda^{'2}_{131}}\cos^{2}\theta_{t}}
\over{4\pi\alpha}}.
\end{equation}
This situation corresponds to the case of
$\mstl$$<$$m_{t}+m_{\sz1}$, $m_{b}+m_{\swl}$.
While for $\mstl$$>$$m_{t}+m_{\sz1}$ or $m_{b}+m_{\swl}$
BR($\stl\rightarrow ed$) will compete with
BR($\stl\rightarrow t\sz1$) or BR($\stl\rightarrow b\swl$).
The analyses for such case will be presented elsewhere \cite{heavy}.

We expect a clear signal of the stop as a sharp peak in
the Bjorken parameter $x$ distribution.
The peak point corresponds to $x={\mstl^2}/{s}$.
In Fig.~8 we show the $x$ distributions at fixed $Q^2$.
As clearly seen from Fig.~8, the lower $Q^2$ cuts would be very
efficient for suppression of the SM background,
since the $s$-channel stop contribution is independent of $Q^2$.

In Fig.~8 we take the beam to be $e^-$.
The event rates depend on not only the RB coupling
strength $\lambda'_{131}$ but also the kind of beam.
This is shown in Fig.~9, where we fixed $x=0.2$ at the peak value.
It is found that the $e^+$ beam is more efficient than
the $e^-$ one to separate the stop signal from the SM background.
This can easily be understood from the structure of the coupling.
While the $e^-$ collides only with sea $\bar{d}$-quarks in
the proton, the $e^+$ collides with valence $d$-quarks.
The difference of the structure function of the proton is naturally
reflected in the cross sections .
We also note that the polarized $e^{-}_{L}$ and $e^{+}_{R}$ beams
could be advantageous to suppress the SM background.
In Fig.~10 we show the $y$($=Q^{2}/sx$) dependence of the asymmetries
defined as
\begin{eqnarray}
C_{R}\equiv
{\frac
{\ddf\sigma(e^{+}_{R})/\ddf x\ddf y - \ddf\sigma(e^{-}_{R})/\ddf x\ddf y}
{\ddf\sigma(e^{+}_{R})/\ddf x\ddf y + \ddf\sigma(e^{-}_{R})/\ddf x\ddf y}
}
\end{eqnarray}
and
\begin{eqnarray}
A_{e^{-}}\equiv
{\frac
{\ddf\sigma(e^{-}_{L})/\ddf x\ddf y - \ddf\sigma(e^{-}_{R})/\ddf x\ddf y}
{\ddf\sigma(e^{-}_{L})/\ddf x\ddf y + \ddf\sigma(e^{-}_{R})/\ddf x\ddf y}
}.
\end{eqnarray}
We find that the polarized $e^{\pm}$ beams will be effective to
identify the stop signal.
  Next we show in Fig.~11 the searchable parameter region at HERA in
($\lambda'_{131}$, $\mstl$) plane.
The shaded-region is experimentally excluded through
the atomic parity violation experiments \cite{BARGER}.
The area inside the solid contour represents the region accessible
at HERA for the production of more than ten signal events
above the SM background
with $Q^{2}$$>$$10^{3}$ $(\gev/c)^2$ in 100pb$^{-1}$ running.
HERA will open up a large window as for the existence of the stop with
RB coupling at mass up to 200 (270) GeV at the coupling of
$\lambda'_{131}\simeq 0.1$ with $e^{-}$ ($e^{+}$) beams.

   It is well known that the similar $x$ peak could be expected
in the leptoquark production at HERA \cite{leptoquark}.
We should point out that the stop with the RB couplings could be
discriminated from most of the leptoquarks by its distinctive
properties ;  1)the $x$ peak originated from the stop would exist
only in the NC (not exist in the CC) process because there is no RB
stop couplings to the neutrinos, 2)the $e^+$ beams are more favorable
than the $e^-$ beams as mentioned above.
One of the leptoquarks ${\widetilde{S}}_{1/2}$ with the charge
$Q=-2/3$ will give same signature with the RB stop,
if the stop has BR($\stl\rightarrow ed$)$\simeq$100\%.
Again we note that this situation corresponds to the case of
$\mstl$$<$$m_{t}+m_{\sz1}$, $m_{b}+m_{\swl}$.
While for $\mstl$$>$$m_{t}+m_{\sz1}$ or $m_{b}+m_{\swl}$
BR($\stl\rightarrow t\sz1$) or BR($\stl\rightarrow b\swl$)
will compete with BR($\stl\rightarrow ed$) and the stop could be
discriminated from the leptoquark
${\widetilde{S}}_{1/2}$ with the charge $Q=-2/3$ \cite{heavy}.

\section{Concluding remarks}
  We have investigated various production processes
of the stop at HERA energies in the framework of the
MSSM and the RBM.
For the stop with $\mstl\nle$50GeV appropriate kinematical cuts
will enable us to see signals well above the background.
The stop produced via R-breaking interactions shows a sharp peak in the
$x$ distribution due to its $s$-channel resonance.

The discovery of the stop would reveal us simultaneously the existence
of the top flavor and the supersymmetry.
This is a really big impact to the present day particle physics.
HERA would be the ideal machine for our stop searches because it
is $ep$ collider with just suitable energy range for our purpose.

\vskip10pt
Thanks are due to a Grant-in-Aid for Scientific Research of the
Ministry of Education, Science and Culture($\sharp 03240277$).

\appendix{}

  The analytic expression for the cross section Eq.(\ref{rbcr})
is given as follows.
\begin{eqnarray}
{{\ddf\sigma}\over{\ddf x\ddf Q^{2}}}[e^{-}_{L,R}]
&=&{{2\pi\alpha^{2}}\over{x^{2}s^{2}}}
   \left[
\sum_{\i =1}^{2}\tui[e^{-}_{L,R}]\xq u\xq +
\sum_{\i =1}^{2}\tuib[e^{-}_{L,R}]\xq \bar{u}\xq \right. \\
&  & \left. +
\sum_{\i =1}^{4}\tdi[e^{-}_{L,R}]\xq d\xq +
\sum_{\i =1}^{4}\tdib[e^{-}_{L,R}]\xq \bar{d}\xq
  \right],
\end{eqnarray}
where
\begin{eqnarray}
\tuo[e^{-}_{L}]&=&{{(Q^2 -sx)^2}\over{Q^4}}
(\ee\eu +\qz\ael\aur)^2,\\
\tuo[e^{-}_{R}]&=&{{(Q^2 -sx)^2}\over{Q^4}}
(\ee\eu +\qz\aer\aul)^2,\\
\tut[e^{-}_{L}]&=&{{s^2 x^2}\over{Q^4}}
(\ee\eu +\qz\ael\aul)^2,\\
\tut[e^{-}_{R}]&=&{{s^2 x^2}\over{Q^4}}
(\ee\eu +\qz\aer\aur)^2,\\
\tuob[e^{-}_{L}]&=&{{s^2 x^2}\over{Q^4}}
(\ee\eu +\qz\ael\aur)^2,\\
\tuob[e^{-}_{R}]&=&{{s^2 x^2}\over{Q^4}}
(\ee\eu +\qz\aer\aul)^2,\\
\tutb[e^{-}_{L}]&=&{{(Q^2 -sx)^2}\over{Q^4}}
(\ee\eu +\qz\ael\aul)^2 ,\\
\tutb[e^{-}_{R}]&=&{{(Q^2 -sx)^2}\over{Q^4}}
(\ee\eu +\qz\aer\aur)^2 ,\\
\tdo[e^{-}_{L,R}]&=&\left.\tuo[e^{-}_{L,R}]\right|_{u \rightarrow d}, \qquad
\tdt[e^{-}_{L,R}]=\left.\tut[e^{-}_{L,R}]\right|_{u \rightarrow d},\\
\tdh[e^{-}_{L}]&=&-\frb(\stl)
    {{(Q^2 -sx)^{2}}\over{Q^2\utg}}
    (\ee\ed +\qz\ael\adr),\\
\tdh[e^{-}_{R}]&=&0,\\
\tdf[e^{-}_{L}]&=&{{1}\over{4}}\frb^{2}(\stl)
    {{(Q^2-sx)^2}\over{\utg^2}},\\
\tdf[e^{-}_{R}]&=&0,\\
\tdob[e^{-}_{L,R}]&=&\left.\tuob[e^{-}_{L,R}]\right|_{u \rightarrow d}, \qquad
\tdtb[e^{-}_{L,R}]=\left.\tutb[e^{-}_{L,R}]\right|_{u \rightarrow d},\\
\tdhb[e^{-}_{L}]&=&-\frb(\stl)
    {{s^2 x^2 (sx-\mstl^2)}\over{Q^2(\stg)}}
    (\ee\ed +\qz\ael\adr),\\
\tdhb[e^{-}_{R}]&=&0,\\
\tdfb[e^{-}_{L}]&=&{{1}\over{4}}\frb^{2}(\stl)
    {{s^2 x^2}\over{\stg}}, \\
\tdfb[e^{-}_{R}]&=&0
\end{eqnarray}
for $e^{-}$ beams.
The formula for $e^{+}$ beams can be obtained by
the following replacement in the above formula for the $e^{-}$ beams ;
\begin{eqnarray}
e^{-}_{L,R}&\rightarrow& e^{+}_{R,L} \\
q&\rightarrow& \bar{q} \\
\bar{q}&\rightarrow& q
\end {eqnarray}
Here, ${e}^{f}$ denote the electromagnetic charge of
matter fermion $f$, and
\begin{eqnarray}
A_{L}^{f} & \equiv &
-{{{T}_{3}^{f}-{e}^{f}\sin^{2}\theta_{W}}
                      \over{\cos\theta_{W}\sin\theta_{W}}}, \\
A_{R}^{f} & \equiv &
{e}^{f}\tan\theta_{W},
\end{eqnarray}
where ${T}_{3}^{f}$ are the third component of isospin and $\theta_{W}$
is the Weinberg angle.

\figure{
Contour of lighter stop mass $\mstl$ (GeV)
in ($A_{t}$, $M_{\st}$) plane. We take
$M_{\st_{L}}$$=$$M_{\st_{R}}$$=$$M_{\st}$ and
($\mt$(GeV), $\mu$(GeV), $\tan\beta$)$=$
(150, -300, 2) for (a) and (150, -300, 12) for (b).}
\figure{
Feynman rules for the lighter stop $\stl$ couplings to
$g$, $\gamma$ and $Z$. }
\figure{
Feynman diagrams for sub-process
$e^{\pm}g\rightarrow
      e^{\pm}\stl\stl^{*}$. }
\figure{
$\mstl$ dependence of decay width
$\Gamma(Z\to\stl\stl^{*})$.
The horizontal line corresponds to the present upper bound for
$\Delta\Gamma_{Z}$ at LEP. }
\figure{
Excluded parameter region
in $(\tht, \mstl)$ by LEP and searchable region at HERA.
Region under the dotted line corresponds to excluded region at TRISTAN
assuming massless photino, which is valid
for $\mstl-m_{\photino}$ $>$ 8GeV.
}
\figure{
Acoplanarity distribution.
Kinematical cuts are $P_{D,\Db}^{T}$ $>$ 2GeV/c and 5GeV/c,
5 $<$ $Q^2$ $<$ 100$(\gev/c)^2$, $\misP$ $>$ 5GeV/c
and $|\sin\theta_{D, \Db}|$ $>$ 0.1. }
\figure{
Feynman diagrams for sub-process
$e^{\pm}q \rightarrow e^{\pm}q$. }
\figure{
$x$ distribution at fixed $Q^2$.
Adopted parameters are $\mstl=200\gev$,
$\lambda'_{131}=0.25$ and $\tht=0.0$. }
\figure{
$y$ distribution at fixed $x$ using the electron and the positron.
$x$ is fixed at 0.2.
Adopted parameters are $\mstl=140\gev$
and $\tht=\pi/4$. }
\figure{
$y$ distribution at fixed $x$ of differential asymmetries
$C_{R}$(a) and $A_{e^{-}}$(b).
$x$ is fixed at 0.2.
Adopted parameters are $\mstl=140\gev$
and $\tht=0$. }
\figure{
Searchable parameter region at HERA in ($\lambda'_{131}$, $\mstl$).
Kinematical cut is $Q^{2} > 10^{3}$$(\gev/c)^2$. }

\end{document}